\documentclass[12pt]{article}
\usepackage{graphicx}
\usepackage{cite}
\usepackage{caption}
\usepackage{subcaption}
\newcommand{\ttl}{\tilde{t}_L}
\newcommand{\ttr}{\tilde{t}_R}

\newcommand{\GeV}{\mathrm{GeV}}
\newcommand{\TeV}{\mathrm{TeV}}

\newcommand\pubnumber{DPF2013-157}
\newcommand\pubdate{\today}

\def\triumf{TRIUMF Theory Group\\
4004 Wesbrook Mall, Vancouver, BC V6T~2A3
}
\def\ubc{ Department of Physics and Astronomy, University of British Columbia,\\
Vancouver, BC V6T~1Z1, Canada
}

\def\Title#1{\begin{center} {\Large #1 } \end{center}}
\def\Author#1{\begin{center}{ \sc #1} \end{center}}
\def\Address#1{\begin{center}{ \it #1} \end{center}}

\newcommand\pubblock{\rightline{\begin{tabular}{l} \pubnumber\\
         \pubdate  \end{tabular}}}
\newenvironment{Abstract}{\begin{quotation}  }{\end{quotation}}
\newenvironment{Presented}{\begin{quotation} \begin{center} 
             PRESENTED AT\end{center}\bigskip 
      \begin{center}\begin{large}}{\end{large}\end{center} \end{quotation}}
\def\Acknowledgments{\bigskip  \bigskip \begin{center} \begin{large}
             \bf ACKNOWLEDGMENTS \end{large}\end{center}}

\def\beq{\begin{equation}}
\def\eeq#1{\label{#1}\end{equation}}
\def\eeqn{\end{equation}}

\def\beqa{\begin{eqnarray}}
\def\eeqa#1{\label{#1}\end{eqnarray}}
\def\eeqan{\end{eqnarray}}

\let\bar=\overbar

\def\Dslash{\not{\hbox{\kern-4pt $D$}}}
\def\dslash{\not{\hbox{\kern-2pt $\del$}}}

\def\msb{{\bar{\ssstyle M \kern -1pt S}}}

\begin{document}
\begin{titlepage}
\pubblock

\vfill
\Title{Charge and Color Breaking Constraints in the Minimal Supersymmetric Standard Model}
\vfill
\Author{ Nikita Blinov$^{(a,b)}$, David Morrissey$^{(a)}$}
\Address{(a) \triumf}
\Address{(b) \ubc}
\vfill
\begin{Abstract}
The scalar potential of the Minimal Supersymmetric Standard Model (MSSM) admits the existence of vacua with non-vanishing expectation values of electrically and color charged fields. If such minima are deep enough, the physical electroweak vacuum is rendered unstable by quantum tunneling. By comparing the lifetime of the electroweak vacuum with the age of the universe, the MSSM parameter space can be constrained. Furthermore, the appearance of charge and color breaking minima associated with the stop sector is strongly correlated with the Higgs mass, which has been recently measured at the Large Hadron Collider. We carry out a metastability analysis in the stop sector of the MSSM, improving upon previous results. We exclude parts of the parameter space allowed by the Higgs mass measurement.
\end{Abstract}
\vfill
\begin{Presented}
DPF 2013\\
The Meeting of the American Physical Society\\
Division of Particles and Fields\\
Santa Cruz, California, August 13--17, 2013\\
\end{Presented}
\vfill
\end{titlepage}
\def\thefootnote{\fnsymbol{footnote}}
\setcounter{footnote}{0}

\section{Introduction}

Supersymmetry predicts the existence of scalar partners for the Standard Model (SM) 
fermions. These scalar fields can stabilize the electroweak scale against large quantum 
corrections. One consequence of these extra degrees of freedom is that the potential can 
develop minima other than the electroweak (EW) vacuum. These minima can have non-zero 
expectation values for charged and colored scalar fields. If these charge and color breaking 
(CCB) minima are global, the EW vacuum is destabilized by quantum tunneling. 
The lifetime of the false vacuum can be computed and compared with the known 
age of the universe $t_0 = 13.8\;\mathrm{Gyr}$ 
allowing us to place constraints on the model 
parameters. 

Vacuum stability in the Minimal Supersymmetric Standard Model (MSSM) 
has been considered before
~\cite{Kusenko:1996jn, Reece:2012gi,Casas:1995pd, Casas:1996zy,Claudson:1983et,LeMouel:2001sf}.
However, the recent discovery of the Higgs boson and measurement of its mass at the Large Hadron Collider (LHC)
~\cite{higgs_lhc} compels a more careful study of this issue. 
The Higgs mass in the MSSM relies on large loop-level corrections to be 
consistent with the experimental value of $\sim 126\;$ GeV. These corrections 
depend primarily on the scalar top (stop) masses and mixings, which enter 
the scalar potential and can lead to the appearance of CCB minima. 
This connection between the Higgs mass and metastability is 
described in Section~\ref{sec:stability}. 

In this study we seek to update and clarify the stability 
and metastability bounds on the parameters in the stop sector of the MSSM.
In particular we investigate the correlation between the Higgs mass and vacuum stability.
We consider only zero-temperature tunneling in order to remain model-independent with 
respect to the thermal history of the universe. In Section~\ref{sec:results}, 
we find that metastability provides an 
important constraint on the MSSM parameter space, which is complementary to the Higgs mass bounds.
These preliminary results are more constraining than expected from previous studies.
We conclude in Section~\ref{sec:conclusion}.
\section{Vacuum Stability in the MSSM\label{sec:stability}}
The Higgs boson mass in the MSSM at one loop is approximately~\cite{Martin:1997ns}
\begin{equation}
m_h^2 \approx m_Z^2 \cos^2 2\beta + \frac{3}{4\pi^2} \frac{m_t^4}{v^2}
\left(
\ln \frac{M_S^2}{m_t^2} + \frac{X_t^2}{M_S^2}\left(1-\frac{X_t^2}{12M_S^2}\right)
\right).
\label{eq:higgsmass}
\end{equation}
The tree-level piece (first term) must be smaller than $m_Z^2$, therefore requiring 
the loop corrections to be large, $\mathcal{O}(30\;\mathrm{GeV})$. One way to achieve this 
is to increase the mean stop mass $M_S=(m_{\tilde{t}_1} m_{\tilde{t}_2})^{1/2}$. However, if 
the theory is to remain technically natural, at least one light stop is required~\cite{Draper:2011aa,Hall:2011aa,Wymant:2012zp}. 
This can be achieved by considering large stop mixing, i.e. values of the mixing parameter $X_t$ 
that maximize the second term in Eq.~\ref{eq:higgsmass}. 
This occurs for $X_t/M_S \approx \pm \sqrt{6}$. 
The stop mixing parameter $X_t = A_t^* - \mu/\tan\beta$ is related to the trilinear terms 
in the scalar potential:
\beq
V_{\mathrm{MSSM}} \supset A_t y_t \ttr^\dagger \ttl H_2^0 - \mu y_t \ttl^\dagger \ttr H_1^0 + \mathrm{ h.c.}
\eeq{eq:pot}
Large mixing $X_t$ typically requires the trilinear stop coupling $A_t$ to be large, which 
can induce the appearance of a minimum with non-zero expectation values of the stops. If this CCB 
minimum is global (which depends on the size of the quadratic couplings, in particular the 
stop soft masses, $m_{Q_3}^2$ and $m_{u_3}^2$), tunneling out of the EW vacuum becomes possible. 

The tunneling rate to the CCB vacuum can be computed using the Callan-Coleman method~\cite{Coleman:1977py, Callan:1977pt}. 
The false vacuum decay rate per unit volume is 
\begin{equation}
\Gamma/V = C \exp(-B/\hbar),
\label{eq:decayrate}
\end{equation}
where $B$ is the Eucledian action of a classical field 
configuration called the bounce. The (meta) stability of the EW vacuum 
then requires
\begin{equation}
\Gamma^{-1} \geq t_0\;\Rightarrow\;B/\hbar \geq 400, 
\label{eq:actionconstraint}
\end{equation}
where the coefficient $C$ in Eq.~\ref{eq:decayrate} can be estimated on 
dimensional grounds to be $\sim (100\;\GeV)^4$\footnote{The constraint of Eq.~\ref{eq:actionconstraint} 
is only logarithmically sensitive to $C$. Larger energy scales associated with the CCB minimum 
can make $C$ larger, making the bound of Eq.~\ref{eq:actionconstraint} more constraining. Thus this choice is conservative.}. 
The decay rate cannot be computed analytically even in the simplest case of a 
single field scalar potential. Numerical algorithms for the computation of the 
bounce with an arbitrary number of fields exist~\cite{bouncefinding}. 
Here we employ the code CosmoTransitions~\cite{Wainwright:2011kj} to compute 
the tunneling rates. We also check these rates using an independent code.

A previous numerical analysis of metastability in the MSSM yielded an empirical 
upper bound on the size of the trilinear term~\cite{Kusenko:1996jn}
\begin{equation}
A_t^2 + 3\mu^2 < 7.5(m_{Q_3}^2 + m_{u_3}^2),
\label{eq:empiricalbound}
\end{equation}
relaxing the frequently used analytic bound\footnote{This bound is neither 
necessary nor sufficient. See Ref.~\cite{Casas:1995pd} for a discussion.}
\begin{equation}
A_t^2 < 3 (m^2_2 + m_{Q_3}^2 + m_{u_3}^2)
\end{equation}
from Ref.~\cite{Claudson:1983et}. The latter bound is implemented in 
the supersymmetric spectrum calculators SuSpect~\cite{Djouadi:2002ze}
 and FeynHiggs~\cite{Heinemeyer:1998yj}. 

In the next section we investigate the 
validity of this numerical bound in combination with the correct Higgs 
mass constraint in the MSSM. 
\section{Preliminary Results\label{sec:results}}
\begin{figure}[htb]
    \begin{center}
        \includegraphics{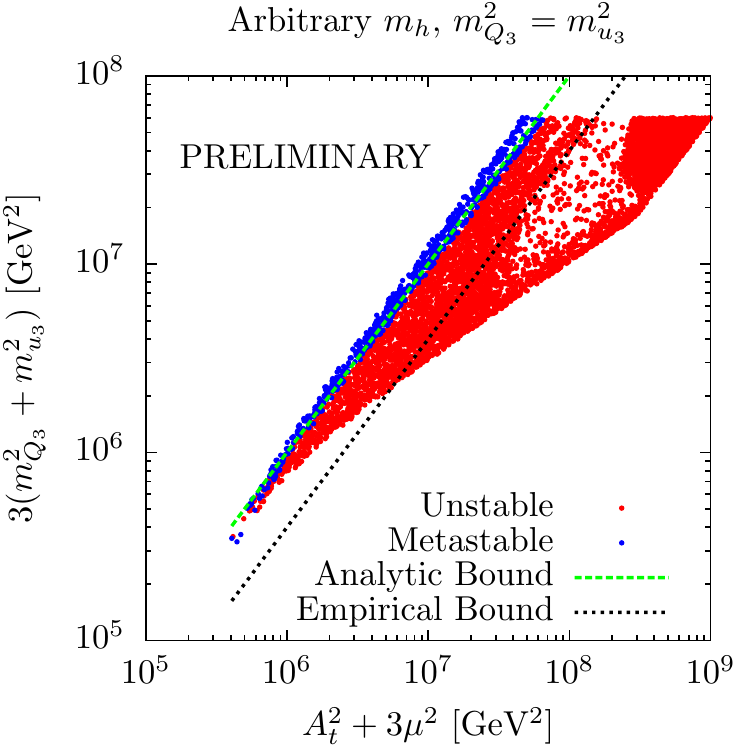}
        \includegraphics[scale=0.9]{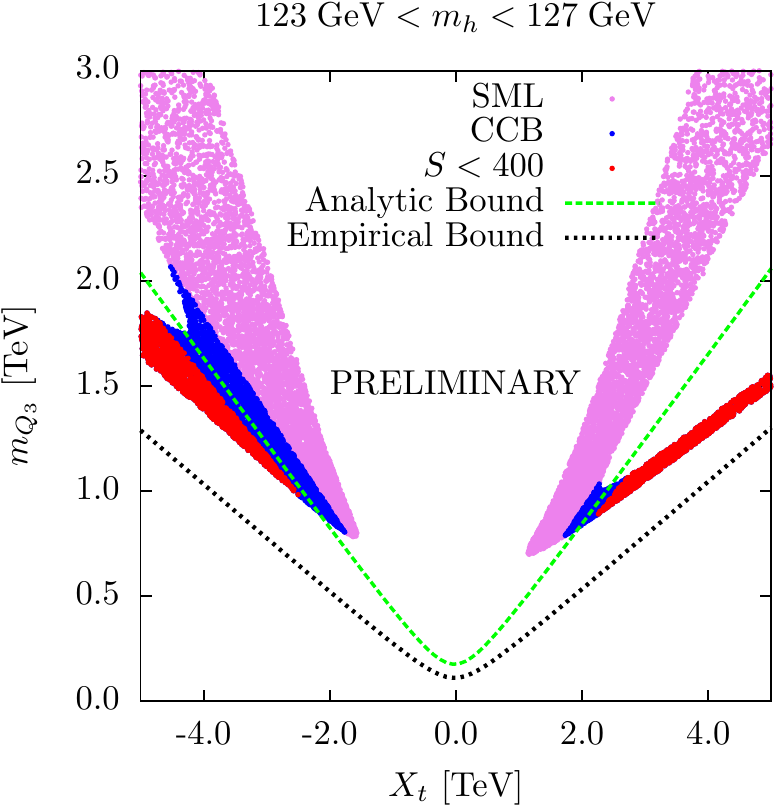}
    \end{center}
    \caption{Metastability in the MSSM. In the left plot we do not impose the Higgs mass constraint. Here every point 
is a model with a global CCB minimum; for red points the tunneling rate out of the SM-like minimum is too fast and these 
are excluded; blue points represent viable metastable points. In the right plot we include the Higgs mass constraint. 
Here blue points are models with a global CCB minimum; red points are excluded due to instability; pink points have an 
absolutely stable SM-like (SML) vacuum. MSSM parameters used in these plots are described in the text.}
    \label{fig:results}
\end{figure}

In order to make the tunneling calculation more tractable we consider 
a restricted field content with only real valued $H_u^0$, $H_d^0$, $\ttl$ and $\ttr$ 
fields. We use the tree-level MSSM potential, taking all parameters to be real. 
To illustrate the general constraints on the MSSM parameter space, we show in Fig.~\ref{fig:results} 
the results for $\tan\beta = 10$, $m_A = 1\;\TeV$ and $\mu=250\;\GeV$; the parameters 
$B\mu$, $m_{H_u}^2$ and $m_{H_d}^2$ are then found by requiring the existence 
of a SM-like (SML) vacuum. We randomly sample $A_t$ and $m_{Q_3}^2$, fixing 
$m_{Q_3}^2 = m_{u_3}^2$, making sure there are no tachyonic sfermions 
in the SML vacuum. For each set of parameters we search for a 
global CCB minimum and, if it exists, compute the tunneling rate 
out of the false SML vacuum. The tunneling rate is then compared 
to the age of the universe using Eq.~\ref{eq:actionconstraint}.

In the left plot of Fig.~\ref{fig:results} 
we show the results of this analysis for an arbitrary Higgs mass. 
Every plotted point represents a model with a global CCB minimum. 
The red points have $B/\hbar < 400$ and are therefore unstable 
and can be excluded. The blue points are viable models, with 
$B/\hbar > 400$ and a metastable SM-like vacuum. For 
small mass squared parameters and large mixings, the mass eigenstates become 
tachyonic, which cuts off the bottom part of the unstable region.  
This plot is a direct comparison with the result of Ref.~\cite{Kusenko:1996jn}. 
We find stronger limits than the empirical bound from Ref.~\cite{Kusenko:1996jn} (black dotted line).

In the right plot of Fig.~\ref{fig:results} we include the Higgs mass constraint. 
We use FeynHiggs to compute $m_h$ at two loops~\cite{Heinemeyer:1998yj}. Here 
we are required to specify the other soft supersymmetry-breaking parameters; 
we take $m_{\tilde{f}} = 2\,\TeV$ and $A_f = 0$ for all sfermions other than the stops, 
and $M_1 = 300\,\GeV$, $M_2 = 600\,\GeV$, and $M_3 = 2000\,\GeV$ 
for the gaugino mass parameters. The colored bands contain 
models with Higgs mass in the range $123\;\GeV< m_h < 127\;\GeV$. 
For large mass squared parameters and small mixings, there is no global CCB minimum 
and the model is absolutely stable; these models are shown in pink. Models with a 
deep CCB minimum are shown in blue and those that are unstable in red. We again 
find that metastability is able to constrain the relevant parameter space, 
contrary to what one would expect from Eq.~\ref{eq:empiricalbound} (black dotted line). 

\section{Conclusion\label{sec:conclusion}}
We have demonstrated that stability of the Standard Model-like ground state in the Minimal Supersymmetric Standard Model 
can provide an important constraint on otherwise viable parameter space. Our results differ significantly from a previous 
numerical study~\cite{Kusenko:1996jn}, showing that metastability is more constraining than expected. 
It is therefore important to ensure that the numerical methods used to compute the tunneling 
rates are under control. We perform these consistency checks in~Ref.~\cite{aterms}.

We have shown results for a particular choice of MSSM parameters. In Ref.~\cite{aterms} we investigate 
the detailed dependence of the metastability limits on these parameters. We also consider flavour and 
electroweak precision observables.
\Acknowledgments
This research is supported by the Natural Sciences and Engineering Research Council of Canada.

\end{document}